# (B)LOCKBOX - Secure Software Architecture with Blockchain Verification

Erik Heiland and Peter Hillmann
Department of Applied Computer Science
Universität der Bundeswehr München
85577, Neubiberg, Germany
E-mail: {erik.heiland, peter.hillmann}@unibw.de

## ABSTRACT

According to experts, one third of all IT vulnerabilities today are due to inadequate software verification. Internal program processes are not sufficiently secured against manipulation by attackers, especially if access has been gained. There is a lack of internal control instances that can monitor and control program flows. Especially when a software vulnerability becomes known, quick action is required, whereby the consequences for an individual application are often not foreseeable.

With our approach (B)LOCKBOX, software building blocks act as verified entities within a transaction-based blockchain network. Source Code, binaries and application execution become supervised. Unwanted interference and manipulation are prevented by the integrity of the distributed system.

## BACKGROUND

To prevent security incidents of manipulated software or transactions in advance, integrated and verifiable solution approaches are required. Our approach focuses on the integrated verification of program flows and communication, which is verifiable by third parties. The goal is to make a dedicated section of program code of arbitrary size verifiable for others in terms of its correctness and freedom from manipulation. In addition, communication between individual nodes of a distributed system can also be managed. (Nagpal 2018, Banerjee et al. 2018)

A civilian example here is payment in a restaurant by credit card. The electronic terminals used for this are often mobile and sometimes change hands. Even if they are subjected to a brief check during development and initialization, manipulation can subsequently take place at various levels and in various ways (Korpela et al. 2017).

Comparable incidents, such as those caused by Stuxnet, can thus be prevented in advance, regardless of which specific vulnerability would be exploited (Falliere et al. 2010).

We address the following research questions:
- How can paying customers be sure that the reader software is trustworthy?
- How can restaurant patrons be sure that the device is not copying their credit card or storing their pin when they pay?
- How do we know that the amount shown on the display of the device also corresponds to the authorized value of the debit?

These questions illustrate only a fraction of the possible attack vectors of critical infrastructures. Especially with the increasing number of IoT devices and the related security issues (Khan and Khaled 2018), it is important to be able to use both trusted software and hardware without having deep expertise in the architecture of each device. Several publications have already investigated the extent to which blockchains can be used to address such security aspects (Conoscenti et al. 2016). Most of them use blockchains only for the immutable storage of data, such as purchases, transaction histories, or for identity management. Nevertheless, such applications are not protected against data being compromised through manipulated hardware or the exploitation of software vulnerabilities.

To our knowledge, there is no approach yet where blockchains are a central part of the software architecture, responsible for both monitoring and controlling the application and ensuring the trustworthiness of the device hardware.

## CONCEPT

The following solution approach using blockchain technology is a preventive countermeasure. Its principle forms the basis for verifying transactions between participants within a network. If this model is applied to the architecture of a software, individual program sections represent the participants (nodes), and transactions represent the function calls, including parameters and return values, between the participants.

Instead of executing a software function or program directly, the calls are routed through so-called guard nodes. These are software systems with special rights that are isolated from the actual application (Deuber et al. 2019). They are responsible for checking the trustworthiness of the network participants and approving communication relationships between them. The process of such a function call and response is visualized in Figure 1.

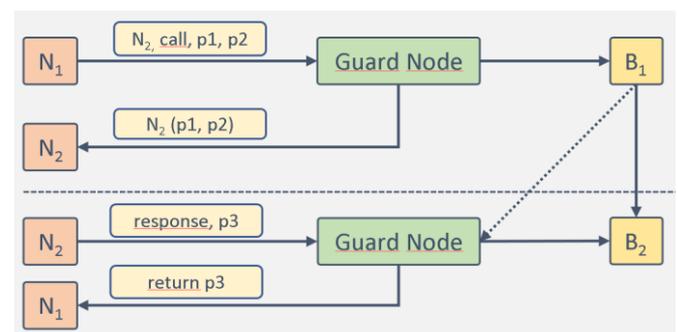

**Figure 1: Transaction between nodes for execution.**

Two different blockchains are distinguished for this purpose. A publicly accessible blockchain (PEBC) manages the software components and hardware devices that have been approved and verified for execution. The second, non-public blockchain (NPBC), is used to store approved program interactions. Both blockchains are managed in a decentralized manner by the guard nodes.

In the PEBC, software applications as well as functions and routines are initially managed for later execution on known systems. The software components are accordingly uniquely identifiable with cryptographic processes and protected against manipulation. Experts analyze this software with regard to correctness and trustworthiness. The data is verified and validated accordingly so that it can then be released for use. Due to the public accessibility for registered participants, verification by third parties can take place here (Skeen 1982). Furthermore, the PEBC also maintains a register of trusted devices. For this purpose, associated public keys are stored for each device. These are used for unique identification, secure communication and signing of data. As soon as software is loaded from a system, the guard nodes check whether it is permitted.

The NPBC is primarily used to store message flows that take place between the nodes of a distributed system. Here, the guard nodes store the communication and program interactions as individual blocks in the decentralized blockchain. In addition, the interaction as well as other inputs and outputs of a user with a node can also be stored. The guard nodes thereby assume a control function for these transactions by notifying the nodes of the permissibility for these as well. Since the transactions can contain sensitive information, they are transmitted encrypted and signed. In addition, the NPBC cannot be viewed publicly.

The data stored in each block is presented in Figure 2. It follows the normal blockchain structure (Hillmann et. al. 2020).

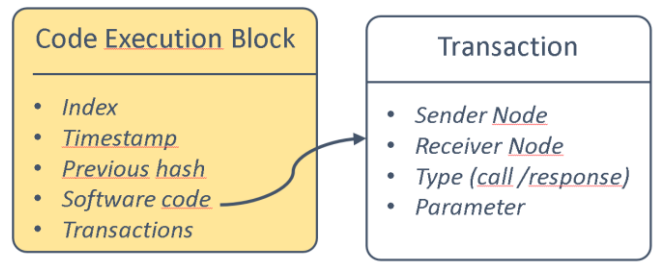

**Figure 2: Structure of the blocks in the blockchain.**

## WORKFLOW

The use of a software component is designed with the (B)LOCKBOX approach as follows. The software to be executed or its individual sections are made verifiable by means of signature procedures. Both the program code or its hash value and the associated signature are stored as initial elements in an PEBC. The decentralized structure means that experts can initially check the code. If the correctness is confirmed in the network, the use of the program code is approved for execution, which is also recorded in the PEBC. If this software is now used on a device, the device itself is first checked via a trusted boot chain. The public keys in the PEBC are also used for this purpose. The desired application is then loaded. This can be done either from the blockchain itself or a copy on the device. Subsequently, the automated verification is carried out by the guard nodes. The executing device or its Trusted Platform Modules (TPM chips) generates a signature for the software that has just been loaded and sends it to the guard nodes for verification. The network of guard nodes performs a comparison with the software applications approved for execution in the PEBC.

In addition to the application functionality, the user can be offered a parallel option for verification, e.g. in the form of a QR code. The verification code is calculated at runtime based on the loaded software. This behavior can be ensured via the well-known technology of the trusted boot chain. Subsequently, the user has the opportunity to convince himself of the trustworthiness of both loaded software and the device. For this purpose, the generated verification codes are matched with the stored signatures in the blockchain.

In addition to ensuring the trustworthiness of software and hardware at program startup, Guard Nodes have another role during program execution. Since all cross-participant function calls are routed through them, an additional security mechanism is created at this point to intercept attack attempts before a routine is executed. Inputs can be checked for plausibility here and authorizations for interaction between nodes can be managed. Only after approval the appropriate function call is passed on and written as block into the NPBC. Such transaction blocks contain the following information: Sender, receiver, transmitted parameters, timestamp and a random number to prevent replay attacks. Each instance of a NPBC represents a complete and unchangeable history of a program flow.

In addition to the additionally gained security barriers within program flows, (B)LOCKBOX creates a complete application log that can be used for debugging, reproduction of specific program states and for integration tests in case of software changes.

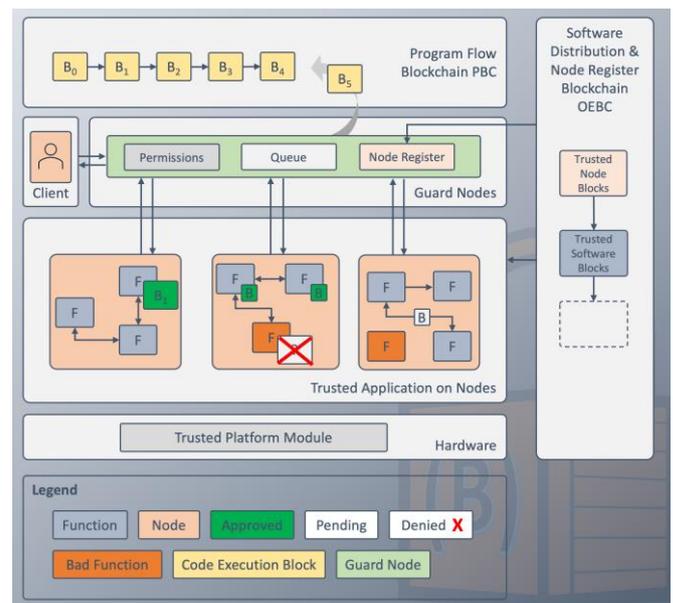

**Figure 3: Overview of the entire system architecture.**

With this solution approach, dynamic procedures can be used securely in addition to static content. This includes the testing of individual function blocks, entire software components, and even complex, distributed systems and their communication processes. Especially when processing information in distributed systems, the transmitted messages and entire transactions can also be checked and tracked.

Figure 3 provide an overview of the main components of the system.

Since all function calls are routed via the guard notes with our approach, a bottleneck may arise here that has an impact on the performance of the application. Load balancing is implemented by instantiating arbitrary numbers of such nodes in the application, which access a common queue in order to be able to check the function calls in a better distributed manner. Another option could be to set up a time-based approval for certain communication relationships between two nodes, in which the data exchange is guaranteed without additional checking by guard nodes. In this form of implementation, however, it must be checked to what extent the blockchain can still be used to reproduce application states, since it is possible that not every call was stored for tracing.

## ASSESSMENT

(B)LOCKBOX can be implemented with currently available technologies for different program types. Both for individual applications and for distributed applications, the security concept can be implemented for different programming languages. A possible structure for the application design is shown as a class diagram in Figure 4.

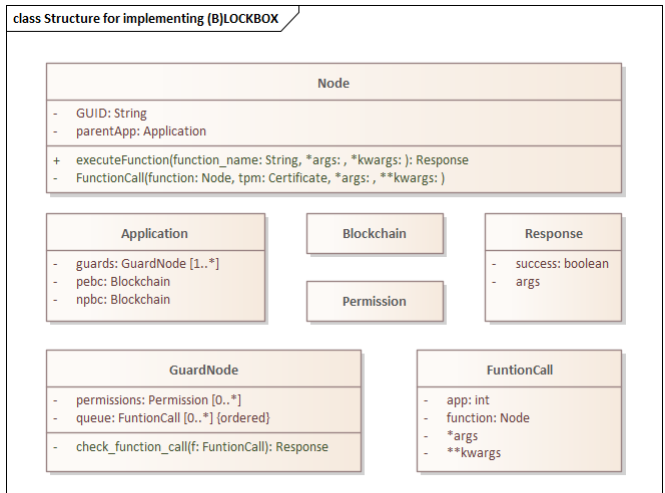

**Figure 4: Possible structure for implementing a (B)LOCKBOX based application**

A node can be a single function or a function group that acts as a self-contained unit and requests all other function calls via its parent application. This application stores the request in the queue so that it can be processed by the next guard node with free capacities. After the permissions have been checked, the transaction is written to the NPBC as a block and the response is written back to the node via the application.

The only public interface of each node is the *executeFunction* function, which is only executed when a Guard Node is passed that was instantiated by the parent application.

Furthermore, the application manages the PEBC and NPBC that are accessed by the Guard Nodes. Each response within the request forwarding contains a truth value, whether the request was approved and the encrypted transfer values.

When applied to existing applications, the challenge is to encapsulate logically and security-related software areas as participant nodes and to redirect function calls via the guard nodes. This requires close cooperation between developers and security experts.

For the technical implementation, existing frameworks for blockchains can be used, so that an implementation does not have to be carried out from scratch. In addition, the individual cryptographic processes used are known, so that they can also be integrated via libraries.

Following the motto "Security by Design", (B)LOCKBOX is best suited for designing new applications. For integration into existing applications, the concept of encapsulated nodes can be set up via a defined interface. Via sandboxing, a large part of the security functionality can already be used. For full integration into existing systems, the internal program processes must be redirected accordingly. A prerequisite for the operation of this architecture are hardware components equipped with TPM chips. This is standard in modern devices, or they can be retrofitted. Based on this, the tried-and-tested principle of trusted boot by means of cryptographic chain is to be applied.

## CONCLUSION

Our approach creates the possibility for multi-level security. Here, targeted requirements can be addressed, and different levels of security and access can be realized.

Known approaches are aimed at preventive measures, whereby runtime aspects are disregarded. Previous methods in this area are software verification and trusted boot, if these are applied. These rely on the trustworthiness of a system. However, since today's systems mainly provide their functionality via a network as a networked system, there are many attack vectors. Current approaches do not allow verification by third parties and do not allow control in and over distributed systems.

With (B)LOCKBOX, both operators and users get easily understandable knowledge about the executed software. Verification gives them control and confidence. In particular, a user can convince himself of the correctness. In addition, depending on the requirements, a verification of the executed inputs, transmitted messages and transactions can be performed, which is a novelty.

Depending on the deployment, a complete application flow can be stored in the blockchain for transactions, resulting in a complete application log. This also enables reproducibility of errors and program states. This allows more detailed analysis and forecasting for the integration of new elements. Using the archived data in the blockchain, meaningful integration tests can be performed.

In summary, the present approach creates an automated system that ensures the correct and trustworthy execution of transactions in distributed systems against multiple cyber attacks.

In the future work we will develop a prototype, which according to our approach realizes a secure software architecture based on a distributed system.

## AUTHOR BIOGRAPHIES

**ERIK HEILAND** is a research associate in the Department of Applied Computer Science at Universität der Bundeswehr München, Germany. He has over 6 years of experience in enterprise architecture modeling and knowledge integration techniques. Currently he is working on his dissertation in the field of ontology alignment. His e-mail address is erik.heiland@unibw.de.

**PETER HILLMANN** is a postdoctoral researcher and scientific supervisor at Universität der Bundeswehr München, Germany. He received a M.Sc. in Information-System-Technology from Dresden University of Technology (2011) and a Dr. rer. nat. (Ph.D. in science) degree in Computer Science (2018) from Universität der Bundeswehr München. His areas of research are system and network security, middleware technologies, operational modeling and optimization. His email address is peter.hillmann@unibw.de.